\documentclass{article}

    \PassOptionsToPackage{numbers, compress}{natbib}

\usepackage[preprint]{neurips_2024}

\usepackage[utf8]{inputenc} 
\usepackage[T1]{fontenc}    
\usepackage{hyperref}       
\usepackage{url}            
\usepackage{booktabs}       
\usepackage{amsfonts}       
\usepackage{nicefrac}       
\usepackage{microtype}      
\usepackage{xcolor}         
\usepackage{graphicx}
\usepackage{amsmath}

\title{Neural encoding with affine feature response transforms}
\author{
    \textbf{Lynn Le\dag}\textsuperscript{1, *}, 
    \textbf{Nils Kimman}\textsuperscript{1, *}, 
    \textbf{Thirza Dado}\textsuperscript{1},
    \textbf{Katja Seeliger}\textsuperscript{2},
    \textbf{Paolo Papale }\textsuperscript{3}, 
    \textbf{Antonio Lozano}\textsuperscript{3}, \\
    \textbf{Pieter Roelfsema }\textsuperscript{3,4,5,6}, 
    \textbf{Marcel van Gerven}\textsuperscript{1}, 
    \textbf{Yağmur G{\"u}{\c{c}}l{\"u}t{\"u}rk}\textsuperscript{1}, 
    \textbf{Umut G{\"u}{\c{c}}l{\"u}}\textsuperscript{1}\\
    \vspace{1em} 
    \begin{tabular}{@{}c@{}}
        \small \textsuperscript{1} Donders Institute for Brain, Cognition and Behaviour, Radboud University, Nijmegen, Netherlands\\
        \small \textsuperscript{2} Max Planck Institute for Human Cognitive and Brain Sciences, Leipzig, Germany\\
        \small \textsuperscript{3} Netherlands Institute for Neuroscience, Amsterdam, Netherlands \textsuperscript{4}  Department of Integrative Neurophysiology,\\
        \small  Centre for Neurogenomics and Cognitive Research, Vrije Universiteit \textsuperscript{5} Laboratory of Visual\\
        \small Brain Therapy, Paris, France \textsuperscript{6} Department of Psychiatry, Amsterdam UMC, University of Amsterdam \\ \\
        \small \textsuperscript{*} These authors contributed equally to this manuscript\\
        \small \dag Correspondence Email: lynn.le@donders.ru.nl
    \end{tabular}
} 

\begin{document}

\maketitle

\begin{abstract}
Current linearizing encoding models that predict neural responses to sensory input typically neglect neuroscience-inspired constraints that could enhance model efficiency and interpretability. To address this, we propose a new method called affine feature response transform (AFRT), which exploits the brain's retinotopic organization. Applying AFRT to encode multi-unit activity in areas V1, V4, and IT of the macaque brain, we demonstrate that AFRT reduces redundant computations and enhances the performance of current linearizing encoding models by segmenting each neuron's receptive field into an affine retinal transform, followed by a localized feature response. Remarkably, by factorizing receptive fields into a sequential affine component with three interpretable parameters (for shifting and scaling) and response components with a small number of feature weights per response, AFRT achieves encoding with orders of magnitude fewer parameters compared to unstructured models. We show that the retinal transform of each neuron's encoding agrees well with the brain's receptive field. Together, these findings suggest that this new subset within spatial transformer network can be instrumental in neural encoding models of naturalistic stimuli. 
\end{abstract}

\section{Introduction}

Elucidating the functional relationship between naturalistic stimuli and their resulting neural responses is a crucial step toward understanding how the brain transforms sensory information into neural representations. A promising approach to this challenge is the development of neural encoding models, which are computational frameworks designed to map sensory inputs to neural responses based on data-driven learning~\cite{vanGerven2017primer}.

One common strategy in neural encoding involves leveraging nonlinear features extracted from deep neural networks trained on categorization tasks. These features are used to encode neural responses in visual areas by linearly mapping the extracted visual features to observed neural activity~\cite{yamins2014performance, gucclu2015deep, st2018feature, khosla2022characterizing, kell2018task, kay2008identifying}. While this linearizing encoding approach has demonstrated potential, it faces several challenges. For instance, estimating large, over-parameterized models from limited data is computationally intensive, as it requires mapping all features in the visual field to all neural response variables. Furthermore, modeling each neural response as a function of all potential spatial stimulus locations complicates the identification of specific spatial computations performed by individual neurons.

Recent research highlights the benefits of incorporating neuroscience-inspired inductive biases into deep neural networks~\cite{kietzmann2019recurrence, banino2018vector, khosla2020neural}. For example, methods such as feature-weighted receptive field (fwRF) models use pre-trained convolutional neural networks to map visual features within spatially localized receptive fields~\cite{st2018feature}. More recent work has sought to decouple "what" and "where" components of neural responses, leveraging deep learning to estimate spatial characteristics and feature tuning simultaneously~\cite{wang2020neural}. This approach refines neural encoding by explicitly addressing the spatial and feature-selective properties of neurons, integrating sparsity and smoothness constraints to provide interpretable receptive field estimates.

In typical linearizing encoding approaches, a naturalistic image containing features of varying complexity is input into a frozen, pretrained convolutional neural network (CNN). As the data propagates through the network's layers, information is systematically extracted to produce visual feature maps. These features are then linearly transformed into singular neural responses. However, unlike this computational process, the biological visual system operates more efficiently. Each neural response in the visual system is spatially specific, responding to particular regions of the visual field rather than the entire image. Research further supports hierarchical increases in receptive field sizes across visual cortical areas and CNN layers~\cite{cadena2019deep}. Moreover, neural responses linked to earlier CNN layers can often be modeled using smaller input tensors. These findings suggest that implementing topographic constraints to selectively propagate relevant features through the network could enhance encoding efficiency.

To address these challenges, we propose the \textit{Affine Feature Response Transform} (AFRT), a novel encoding method designed to minimize redundant computations and enhance both interpretability and precision. This approach models each neuron’s receptive field as an affine retinal transform followed by a localized feature response. The affine transform accounts for distortions in the neural response’s sensory input by learning the contiguous visual field region encoded by multi-unit activity (MUA) signals. By capturing the spatial organization of afferent inputs, this framework reflects the spatial specificity of individual neural responses and is adaptable to arbitrary neural data~\cite{hubel1962receptive}.

The AFRT approach achieves the following key advancements:
\begin{enumerate}
    \item It introduces a novel subset of spatial transformer networks (STNs)~\cite{jaderberg2015spatial}, tailored for neural encoding tasks.
    \item By employing only three learnable parameters—shift ($x$, $y$) and scale ($s$)—AFRT significantly reduces the number of parameters required compared to unstructured models.
    \item Incorporating anatomically grounded inductive biases enables the encoding of MUA in visual cortical areas V1, V4, and IT of macaques, leading to improved performance over state-of-the-art linearizing encoding models.
    \item The method enhances interpretability by visualizing each neuron’s encoding through its learned retinal transform.
    \item Interestingly, the model reveals that larger receptive fields are needed to predict neural responses in V4 and IT compared to V1, offering new insights into retinotopic mapping.
\end{enumerate}

By introducing AFRT, we provide a robust framework for advancing our understanding of how the brain encodes sensory inputs, paving the way for more efficient and interpretable neural encoding methods.


\section{Methods}
\subsection{Preliminaries}
We aim to develop a model, $f_{\theta}$, that predicts neural responses $r \in \mathbb{R}$ from sensory stimuli $s \in \mathbb{R}^{H \times W \times C}$, denoting:
$$f_{\theta}: s \mapsto r$$
Here, $\theta$ represents the parameters of a neural network.

Traditionally, the model $f_{\theta}$ is constructed using a convolutional network pre-trained for object recognition, attached to a learned linear transformation $w_\text{global} \in \mathbb{R}^{D \times H' \times W'}$~\cite{yamins2014performance, gucclu2015deep}. The response is then computed as:
$$f_{\theta}(s) = w_\text{global}^\top \phi(s)$$
where $\phi(s) = z \in \mathbb{R}^{D \times H' \times W'}$ is the output of the feature extractor, mapping the input $s$ to features $z$. The linear transformation by $w_\text{global}$ pools features across the entire feature space to produce the response $r$.

This setup, while expressive, lacks structural inductive biases that could improve generalization and interpretability. To address this, we introduce a spatial transformation within the feature extraction process:

$$f_{\theta}(s) = w_\text{local}^\top \phi(T(s))$$

Here, $T$ represents a spatial transformer network~\cite{jaderberg2015spatial} that modifies the input $s$ via a constrained affine transformation $A \in \mathbb{R}^{2 \times 3}$. While a general affine transformation typically involves six parameters (rotation, scaling, shearing, and translation), our implementation is constrained to three parameters: two translations $(t_x, t_y)$ and one scaling factor $s$. The transformation adjusts $s$ before feature extraction $\phi$, ensuring that $\phi$ operates on the transformed input $T(s)$, and $w_\text{local}$ then aggregates these features into $r$.

By using this simplified transformation, the model prioritizes preserving key spatial relationships such as scaling and translation invariance while reducing the risk of geometric distortions introduced by more complex transformations like rotation or shearing. This constrained parameterization enhances both interpretability and parameter efficiency while maintaining sufficient flexibility to align input stimuli spatially.

\subsubsection{Affine feature response transforms}
The core principle of AFRT is to model each neuron's receptive field as a sequential process comprising an affine transformation followed by localized feature extraction. This approach differs from the unstructured model in its weight space: AFRT utilizes spatially constrained weights $w_\text{local}$, whereas the unstructured model relies on global weights $w_\text{global}$.

Each neuron's response at a spatial position $(x, y)$ is modeled as:
$$r = w_\text{local}^\top \phi(T_{\theta}(s; x, y))$$
where $T_{\theta}$ is a spatial transformer network parameterized by $\theta$ that produces the transformed input $V = T_{\theta}(s; x, y)$. The affine transformation $A$ aligns the input $s$ so that pixel $(x, y)$ is centered through translation and rotation, producing the spatially adjusted input $T_{\theta}(s; x, y) = A s$. The pre-trained convolutional feature extractor $\phi$ then operates on this adjusted input, and the localized weights $w_\text{local}$ act on the output channels of $\phi$.

This approach decouples the affine transformation $A$ from the complex feature extraction $\phi$, enhancing both interpretability and parameter efficiency. By decomposing each neuron's response into spatial alignment and feature extraction, AFRT provides a structured framework for understanding how sensory inputs are processed and encoded by neural mechanisms.

\subsection{Model optimization}
AFRT models are optimized using datasets $D = \{(s_i, r_i)\}$ of recorded neural responses by minimizing the mean squared error (MSE) loss:
$$L(\theta, w_\text{local}; D) = \sum_i (r_i - f_{\theta}(s_i; w_\text{local}))^2$$
where $f_{\theta}(s_i; w_\text{local}) = w_\text{local}^\top \phi(T_{\theta}(s_i; x_i, y_i))$ represents the predicted response for the $i$-th recording, focused on point $(x_i, y_i)$.

During optimization, the weights $w_\text{local}$ and the parameters of the affine transformation $A$ are learned, while the pre-trained features from $\phi$ remain fixed. The model leverages the Adam optimizer~\cite{kinga2015method} to refine both $w_\text{local}$ and $A$. This structured optimization ensures that the model adapts spatial transformations while maintaining computational efficiency.

The AFRT framework accommodates standard convolutional architectures, such as VGG~\cite{simonyan2014very} or ResNet~\cite{he2016deep}, and enhances their utility within the structured context of affine transformations and disentangled weights. Next, we examine the advantages brought by AFRT in terms of interpretability and parameter efficiency.

\subsection{Theoretical justification}
We now analyze how AFRT’s structure provides benefits in terms of parameter efficiency, optimization, and generalization compared to unstructured models.


\paragraph{Parameter efficiency}
A key advantage of AFRT is its parameter efficiency. Each neuron’s encoding is factorized into an affine warp $\mathcal{A}$ and local feature weights $w$. For a neuron with receptive field size $\mathcal{R} \times \mathcal{R}$ pixels on an input of $W \times H$ pixels, this requires estimation of 
$\mathcal{A}$, consisting of three parameters (2 translation, 1 scale) and estimation of  $w$, consisting of $ \mathcal{O}(\mathcal{R}^{2} \cdot C)$ parameters, where $C$ is number of feature channels
In contrast, an unstructured model requires $\mathcal{O}(W \cdot H \cdot C)$ parameters to linearly combine global features. For example, on $224 \times 224$ images with $c=128$ channels and $7 \times 7$ receptive fields, AFRT requires $3 + 49 \cdot 128 = 6275$ parameters per neuron compared to $128 \cdot 224 \cdot 224 = 6,423,552$ for the unstructured model -- a three order of magnitude difference. This massive reduction in parameters helps regularization and generalization, as the encoding model is less likely to overfit.

\paragraph{Modeling low-order dependencies}
Learning unstructured linear weights \(w\) over features \(Z\) not only presents optimization challenges and risks of overfitting due to the high dimensionality of \(w\), but it also poses a high risk of overfitting due to complex higher-order dependencies among the features \(Z\). In the AFRT model, the features \(Z = \phi(T_{\theta}(S))\), transformed by \(\mathcal{A}\), are spatially clustered according to each neuron's receptive field. Consequently, \(w\) primarily needs to model simpler, low-order dependencies within these localized regions of \(Z\). Together, this simpler optimization landscape enables efficient training and better generalization compared to highly underconstrained unstructured models.

\paragraph{Generalization to natural settings}
By factorizing spatial transformations from feature computation, AFRT adheres more closely to the anatomy of biological vision systems. Sensor measurements are rectified into local coordinate frames through mechanical and neural feedback control \cite{gilbert2013top}. Downstream feature tuning is thus inherently local. Unstructured models lack this inductive bias, instead learning complex globally entangled weights. AFRT’s biological realism can thus improve generalization to natural settings. 

In summary, AFRT’s structure confers substantial benefits in terms of parameter efficiency, trainability, generalization, and biological fidelity. 

\subsection{Experiments}
\label{sec:experiments}

Next we describe the empirical validation of these advantages.

\paragraph{Experimental data}
The THINGS database is well known for its high amount of naturalistic object images (26,000) and diverse object concepts (1,854). There were a total of 25,248 natural images presented to the monkey from the THINGS image database; 12 images of each of the 1,854 stimulus categories were used. A detailed description of the THINGS database \footnote{https://things-initiative.org} is provided in Hebart et al.~\cite{hebart2019things}. 

A macaque monkey was implanted with 1024-channel implant consisting of 16 Utah arrays~\cite{chen2020shape, chen20221024}. 7 arrays are placed in the V1, 4 in the V4, and 4 in the IT. The 25,248 images were divided into a training and a test set, and presented randomly and interleaved. The training set contained 12 images per category, which were shown once. The test set comprised 100 images that were shown 30 times each. The monkey fixated for 300 ms on a red dot with a gray background and then a fast sequence of 4 images were shown, with 200 ms of presented stimuli and 200 ms inter trial interval. The shown images contained 500 by 500 pixels, and were shifted to the lower-right fovea by 100 pixels in the x and y axis. If fixation was kept for all the sequence, the monkey got juice reward.  

The recorded multi-unit activity (MUA) responses are extracellular signals from local neuron networks, believed to represent the collective spiking activity of these neurons \cite{burns2010comparisons}. Initially, the raw data was averaged over time to reduce noise. Subsequent normalization involved subtracting the mean response of all test trials for that day from each individual trial and channel, followed by division by the standard deviation of these trials. To assess the reliability of the data, correlations were computed for all possible pairs of the 100 test images, resulting in 435 Pearson correlation coefficients per electrode channel \((30 \times (30-1) / 2)\). These correlations serve as reliability scores used to threshold the data, ensuring consistent analysis across trials and channels. A total of 1024 recording sites provided 1024 neuronal signals, and after filtering out signals with a reliability score of lower than 0.4, we were left with 667 electrode channels. 


\begin{figure}
\includegraphics[width=\textwidth]{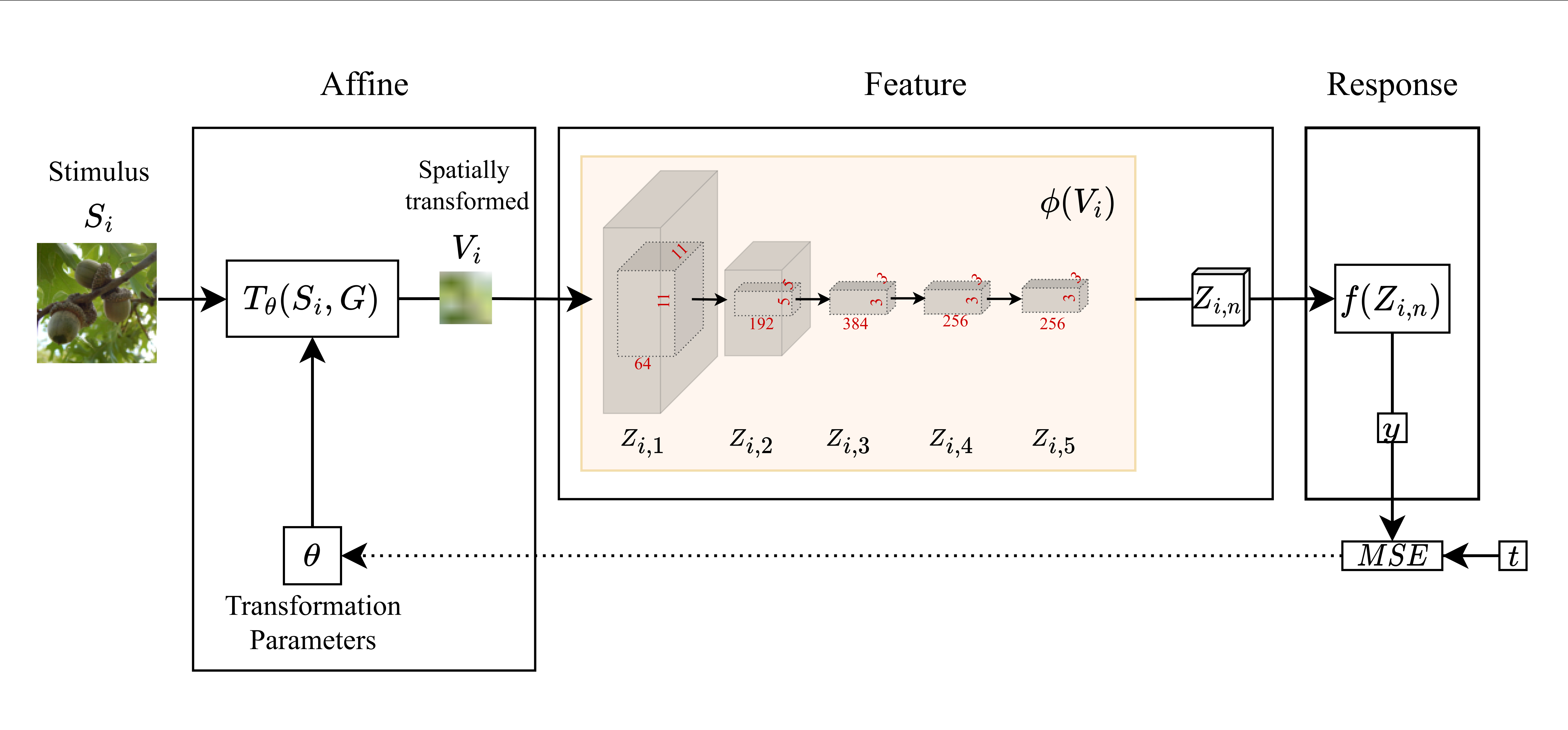}
\caption{ A schematic overview showing the training procedure of the AFRT model. The input images are passed through the affine module. The image becomes scaled and cropped based on $\theta$. The resulting spatially transformed images $V_i$ are then passed through the feature model $\phi(V_i)$. Then the response layer $f(Z_{i,n})$ converts the features into predicted responses $y$.}\label{fig:fig1}
\end{figure}

\paragraph{Models}

We trained our AFRT model on the neural dataset. Our full affine feature response transform model applies a spatial transformation on the stimulus ($S_i$) with learnable parameters $\theta$ to produce an intermediate representation $V$, which is the affine-transformed input (see Fig. \ref{fig:fig1}). The dimensionality of $V$ is scaled to either 16, 32, or 64, tailored to the specific feature layer involved in the encoding process (refer to Table \ref{tab:tab1}). This scaling ensures that $V$ remains compact, optimizing it for efficient processing through the network and minimizing computational overhead. The feature extractor $\phi$, which is a pretrained AlexNet on ImageNet, processes $V$ to extract features from five convolutional layers: Conv1 ($Z_{1}$), Conv2 ($Z_{2}$), Conv3 ($Z_{3}$), Conv4 ($Z_{4}$), and Conv5 ($Z_{5}$). These features are then linearly transformed to generate the final neural response.

\begin{table}[!ht]
  \caption{Value of the field of view parameter per layer of AlexNet examined. Deeper layers are more complex and can accept higher resolution input images.}
  \label{tab:tab1}
  \centering
  \begin{tabular}{lll}
    \toprule
    Identifier & $V$ size \\
    \midrule
    $Z_{1}$ & 16  \\
    $Z_{2}$ & 32  \\
    $Z_{3}$ & 32  \\
    $Z_{4}$ & 32  \\
    $Z_{5}$ & 64  \\
    \bottomrule
  \end{tabular}
\end{table}

As a baseline, we trained an unstructured linear model that utilizes features extracted from a pretrained AlexNet. This model follows the standard neural encoding approach described in prior work, such as G\"u\c{c}l\"u et al.~\cite{gucclu2015deep}, where features from a pretrained convolutional network are used to predict neural responses. Unlike AFRT, this model does not incorporate affine transformations and therefore does not account for specific feature locations within the input image. Instead, all input stimuli ($S$) are processed at a uniform size of $224 \times 224$ pixels, consistent with the default input size of AlexNet. This ensures consistency across all feature layers and neuronal signals but does not leverage spatial specificity.

\paragraph{Training parameters}
\label{sec:hyperparams}
For our AFRT model, the affine warps \(\mathcal{A}\) were initialized to identity. The dataset was divided into 22,348 training samples and 100 test samples. The linear weights $w$ were initialized to uniform average pooling. Each model is trained 100 epochs with a batch size of 100 samples. All models were optimized using Adam with learning rate 0.0002 and batch size 4 and 100 epochs. The architectures and training loops are implemented with the Mxnet library~\cite{chen2015mxnet}. The source code and detailed implementation can be found in our repository \footnote{\url{https://github.com/lelynn/AFRT}}. 

\subsection{Performance evaluation}
To evaluate performance, we trained three encoding models for each MUA signal, using features from layers 1, 2, and 5. For each MUA, we selected the best-performing model out of the three trained models based on the Pearson correlation value between the predicted response and the target response. This method not only provides a large space of models to select from but also identifies which layer contains the most informative features for the encoding task.

The receptive field (RF) size corresponds to the size of the result of the affine transformation applied to the original image. Specifically, this transformed region determines the portion of the original image contributing to the neural response at the layer being analyzed. By using the parameters of the best-performing model, we identified the effective RF size in the original image space, ensuring consistency with the spatial transformations and feature extraction applied during analysis.

\section{Results}

\subsection{Accuracy of MUA predictions}

\begin{figure}[!ht]
\centering
\includegraphics[width=0.8\textwidth]{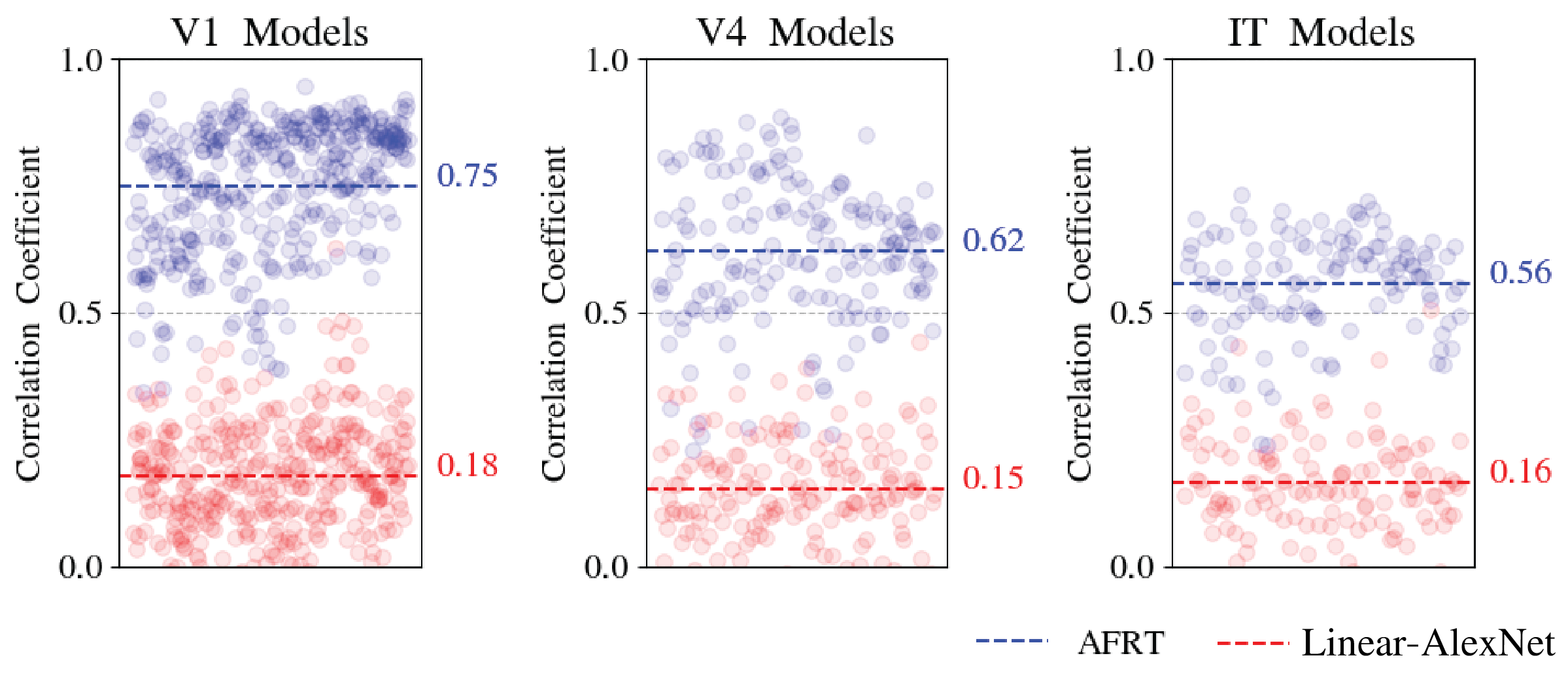}
\caption{Comparison between the performance values of the AFRT model (blue) and the baseline model (red). Single blue dots show correlation values for trained AFRT models trained and red dots show the values of baseline models trained. The dashed line show the average across all models. Both models are trained on the training set and values are evaluated using the test set. The top row shows all the models that were trained using three feature layers per electrode (1122 models for V1, 507 for V4, and 372 for IT) whereas the bottom row shows the best selected performance (374 models for V1, 169 for V4, and 124 for IT).}
\label{fig:fig2}
\end{figure}

Our analysis shows that the predicted activity from the AFRT model correlate higher with ground truth signals compared to the predicted activity from the baseline model Linear-AlexNet (Fig. \ref{fig:fig2}). We plotted the correlation values for all the best performing models from conv1, conv2 and conv3 layers. Each point represents a signal-wise model, and the color represents the model type (blue is AFRT, red is Linear-AlexNet). Although the baseline model makes use of a significant higher amount of features, our results show that models containing the affine components perform better (with a correlation value of 0.5 or higher). Overall, AFRT encodes MUA activity more accurately than the Linear-AlexNet model and our results also show that AFRT is less prone to overfitting.

\subsection{Models of downstream brain regions contain larger receptive fields}

The feature model retains constant features while learning only the affine parameters and the response layer. This process aligns the input features with those of the feature model, revealing specific regions within the visual space that provide optimal information for effectively encoding the MUA response.

To assess whether the AFRT model captures realistic retinotopic properties while predicting MUA responses, we visualized the transformed input selections determined by the AFRT for each model as squares on a plot. The constraints on these transformation parameters ensure alignment within the visual field; however, the resultant square locations are not directly comparable with actual retinotopic data. It is crucial to recognize that MUAs cover only a limited portion of the visual field, attributed to the invasive nature of the recording. Consequently, we focused on comparing the sizes of these transformed selections across different brain regions.

\begin{figure}[!ht]
\includegraphics[width=\linewidth]{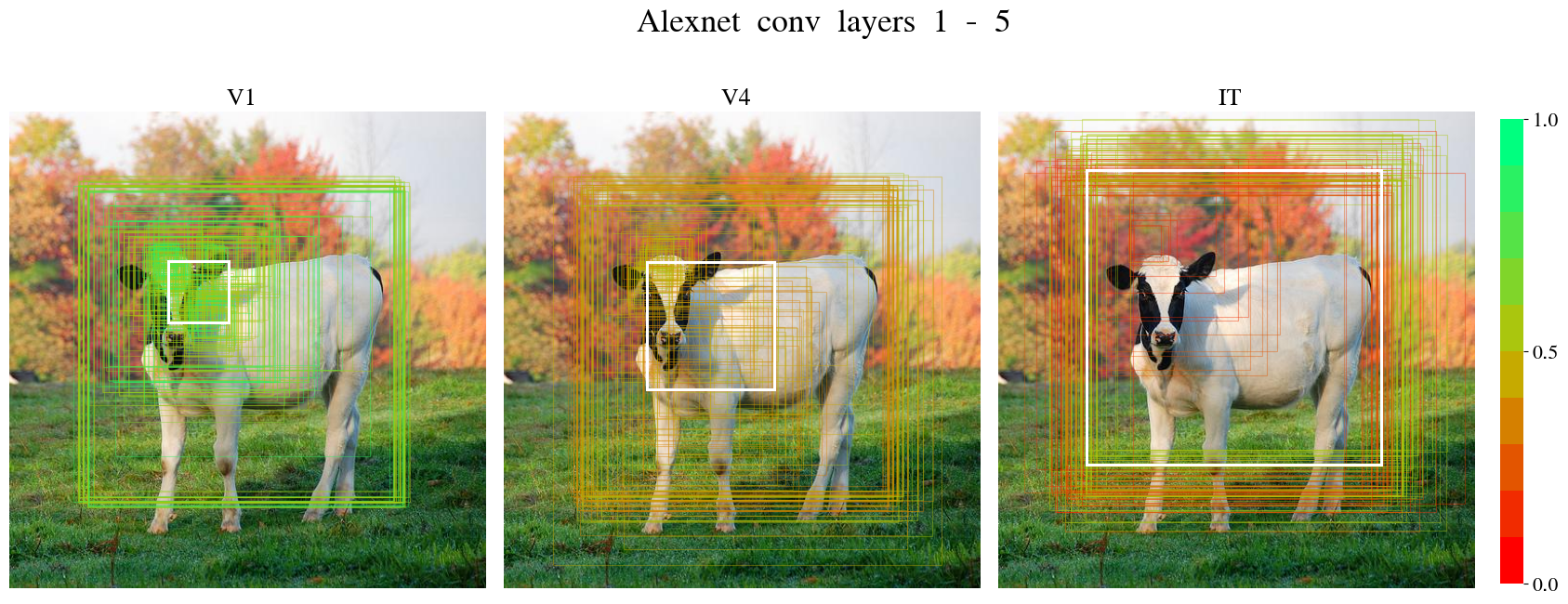}
\caption{Receptive fields of all the best performing models, separated by brain
region. The colored squares are individual receptive fields, colors are indicative of model performance (Pearson R correlation on the test set). The white squares indicate the squares made from averaged learned $\Theta$. Note that the amount of models over these regions vary: V1 has twice as many models (374) as V4 and IT (169 and 123 respectively).}
\label{fig:fig3}
\end{figure}

In Fig.~\ref{fig:fig3}, we observe that deeper brain regions exhibit larger transformation selections, consistent with established principles regarding the scaling of neuronal receptive fields. The variance observed in the sizes of the AFRT-learned receptive fields (RFs) across various models suggests a model preference for specific locations extracted from the input for encoding purposes.
Notably, some receptive fields in V1 are quite large, potentially reflecting the nature of MUA signals, which may aggregate information from multiple neurons.

\subsection{Downstream brain regions are better encoded from deeper AlexNet features}

\begin{figure}[!ht]
    \centering
    \includegraphics[width=0.5\linewidth]{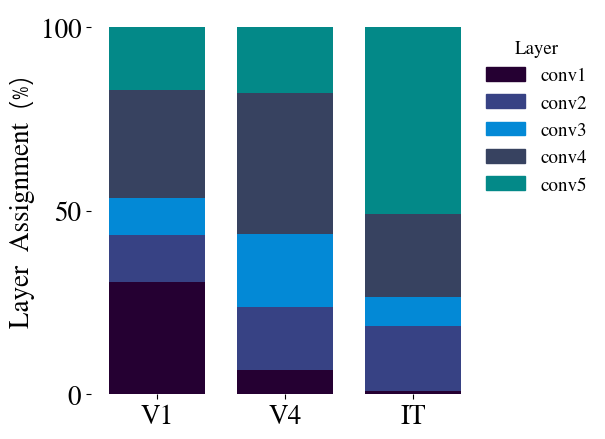}
    \caption{Layer contribution to model performance averaged over amount of signals in V1, V2, and V4. Each column shows the layer contribution to the model's performance for a single ROI. The colors indicate the \% of contribution each DNN layer had over all signals in that ROI.}
    \label{fig:fig4}
\end{figure}

MUA signals were systematically categorized based on their respective brain regions, revealing a progressive shift towards higher layer assignments when moving from V1 to V4 in the visual cortex of the macaque (Fig.~\ref{fig:fig4}). To facilitate comparisons between brain regions, despite variations in the number of models per region, the data was normalized to 100\%. The objective was to identify the AlexNet layer that provided optimal encoding performance for neurons across different brain regions.

The findings corroborate our initial hypothesis: earlier AlexNet layers tend to better model the neural activity in earlier visual cortex regions, with this pattern persisting for deeper layers corresponding to more advanced brain regions. This observation is consistent with previous research by G\"u\c{c}l\"u et al.~\cite{gucclu2015deep}, who noted that some higher convolutional layers of VGG and AlexNet exhibit Gabor-like features, likely providing a good fit for the initial visual processing stages.

In each brain region, about 50\% of the models preferred the layer that yielded the best results, whereas the other half selected different layers, with less significant contributions. An interesting anomaly occurs in V1, where despite layer five's strong performance, it does not match the effectiveness of layer one. This could suggest that complex features from layer four, when applied to the simpler neural structures of V1, might lead to overfitting.




\section{Discussion}
In this study, we introduced the affine feature response transform, a novel adaptation of existing linearizing encoding models \cite{naselaris2015voxel, gucclu2015deep, yamins2014performance}. The AFRT model integrates retinotopic mapping as a core hypothesis, assigning each neuronal response to a specific visual field location while minimizing redundancy through three primary image transformation parameters: shift (x,y) and scale. This approach aligns with recent findings where the integration of biologically-inspired components into neural encoding models has enhanced fMRI prediction accuracy \cite{khosla2020neural, st2018feature}.

Our model demonstrates substantial enhancements in predicting multi-unit activity (MUA) across the V1, V4, and IT regions of the macaque, outperforming traditional models that lack biologically-inspired constraints. Additionally, AFRT significantly reduces the number of required parameters by transforming feature responses into scalars instead of entire feature maps, as illustrated in Figure \ref{fig:feature_shapes}. This reduction not only simplifies the model complexity but also improves the interpretability and efficiency of response predictions. Furthermore, while our study employs basic assumptions—that each neural signal corresponds to a non-rotating spatial receptive field—the proposed AFRT model is not inherently limited to these constraints. Indeed, spatial transformer networks could extend this model by incorporating additional parameters, allowing for rotation of the spatially transformed image \cite{jaderberg2015spatial}. 

In our model, each neural response is associated with a specific spatial location within the visual field. Adapting the model to accommodate multiple receptive fields per neural response could significantly enhance its applicability, particularly because recording sites often contain signals from multiple neurons. Additionally, integrating temporal dynamics and movie data, as opposed to solely static images, might reveal whether and how spatial receptive fields vary under dynamic conditions. Our study shows first effort to assign each model to a spatial location in the input with the aim to significantly reduce the number of learnable parameters while enhancing accuracy of the encoding model. We utilized a pretrained AlexNet model on ImageNet as a feature extractor within our framework. Future work could involve training this model end-to-end to potentially improve performance and elucidate features, aligning with findings that suggest data-driven training of encoding models could enhance the prediction of macaque V1 responses to natural images \cite{cadena2019deep}.

Beyond enhancing our understanding of visual processes, neural encoding models also hold potential for applied domains. For instance, these models can facilitate advancements in cortical prosthetics, potentially improving the accuracy of prosthetic virtual reality simulations that aim to stimulate visual perceptions with greater accuracy \citep{vanSteveninck2022end}.

\subsection{Broader impact} \label{sec:broaderimpacts}
Neural encoding models, particularly those designed to predict neural activity from naturalistic images, significantly enhance our understanding of how visual stimuli are processed and represented in the brain. These models, incorporating aspects of retinotopy, are pivotal in elucidating the complex relationship between external visual environments and their corresponding neural responses. Such models are crucial for developing advanced visual neuroprosthetics aimed at simulating neural activity to possibly restore vision. Nevertheless, the application of these models must be undertaken with prudence, given the intricate nature of brain functionality and its interaction with the environment.

Human interaction with surroundings transcends mere visual reception and involves intricate behaviors and neuroplastic changes that models based solely on retinal inputs might not fully address. For instance, the dynamic nature of visual processing in response to moving stimuli and the resultant motor behaviors add layers of complexity not typically modeled by static visual inputs. Additionally, employing these models for predicting the neural impact of visual stimuli in neuroprosthetic devices may not completely mimic the natural experiences due to differences in how eyes are fixated in experimental set-up.

Moreover, insights gained from neural encoding models could revolutionize how we understand and enhance cognitive engagement with visual stimuli, potentially improving educational and therapeutic strategies. However, the advancement of such technologies also poses ethical risks, particularly if used to infer personal or sensitive information without consent. Although the practical misuse of this technology remains limited—owing largely to the complexities of accurately modeling individual neuronal patterns—the ethical considerations are significant and must be vigilantly evaluated as the technology progresses.

\subsection{Limitations}
The constraints on transformations enhance interpretability by focusing on biologically plausible manipulations, such as scaling and translation, but they also limit the range of neural dynamics the model can capture. For instance, real neural receptive fields may involve rotation or shear under specific conditions, such as attention or learning, which are not modeled here.

By excluding rotation and shearing, the constrained affine transformations preserve parallelism and proportional scaling, simplifying the parameter space and making the learned adjustments more interpretable. However, this trade-off may restrict the model's ability to represent neural responses dependent on more complex geometric properties. Further investigation is needed to explore how such constraints influence neural encoding.

Additionally, the applicability of these findings to non-invasive imaging techniques, such as functional magnetic resonance imaging (fMRI), remains unclear. Unlike invasive methods with high spatial resolution, fMRI has a lower signal-to-noise ratio and lacks the fine-grained detail provided by electrode arrays. This distinction is significant, as invasive methods are rarely performed on human subjects.
\label{sec:limitations}

{
\small
\bibliographystyle{unsrt}

}

\appendix

\section{Feature shapes example}
\begin{figure}[!h]
\includegraphics[width=\linewidth]{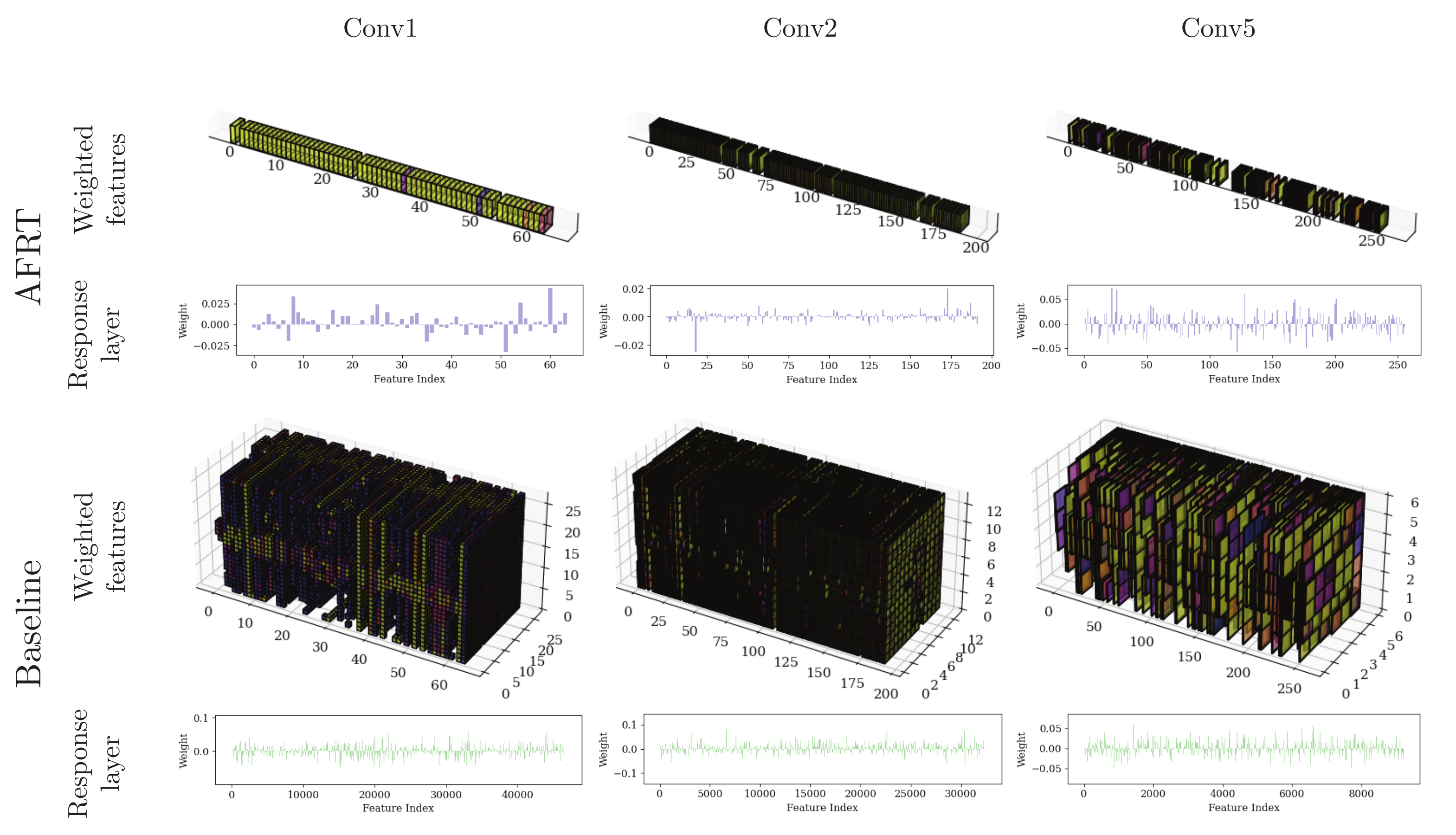}
\caption{Example of weighted features from AFRT and baseline. This is three example layers from 5 trained layers, of one model.}
\label{fig:feature_shapes}
\end{figure}

In Fig.~\ref{fig:feature_shapes} we show the dimensional structure of feature spaces that are then transformed by the linear models into scalar responses. Specifically, for AFRT, the features are represented simply as \((\text{depth}, 1, 1)\), indicating a singular, depth-wise vector per feature. In contrast, the regular linearizing-AlexNet encoder contains a considerably larger feature space for each layer.

\end{document}